\documentclass[prl,preprintnumbers]{revtex4}
\usepackage{dcolumn}
\usepackage{bm}
\usepackage{ulem}
\usepackage{epsfig}
\usepackage{epstopdf}
\usepackage{amsmath}
\usepackage{graphicx}
\usepackage{mathrsfs}
\usepackage{amssymb}
\usepackage{amsfonts}
\usepackage{hyperref}
\usepackage{color,soul}
\usepackage{float}
\usepackage{enumerate}
\usepackage{subfigure}
\usepackage{xcolor}
\usepackage{amsthm}
\usepackage{braket}
\usepackage{multirow}    
\usepackage{array}     

\begin{document}

\title{An experimental measurement of the topological charge of  orbital angular momentum beams through weak measurement}


\author{Jing Zhu,$^{1,2}$ Pei Zhang,$^{1 *}$  Qichang Li,$^1$  Feiran Wang,$^1$ Chenhui Wang,$^1$ Yingnan Zhou,$^1$ Jinwen Wang,$^1$ Hong Gao,$^1$, L. C. Kwek$^{2 *}$, and Fuli Li$^1$}

\affiliation{
$^1$Shaanxi Key Laboratory of Quantum Information and Quantum Optoelectronic Devices, Xi'an Jiaotong University, Xi'an 710049, China\\
$^2$Center for Quantum Technologies, National University of Singapore, Singapore 117543\\
$^*$Corresponding author: zhangpei@mail.ustc.edu.cn, kwekleongchuan@nus.edu.sg.
}

\begin{abstract}

As a special experimental technique, weak measurement extracts very little information about the measured system and will not cause the measured state collapse. When coupling the orbital angular momentum (OAM) state with a well-defined pre-selected and post-selected system of a weak measurement process, there is an indirect coupling between position and topological charge (TC) of OAM state. Based on these ideas, we  propose an experimental scheme that experimentally measure the TC of OAM beams from -14 to 14 through weak measurement.  

\end{abstract}
\maketitle


Beams with an azimuthal phase profile of the form $\exp(il\varphi)$ may carry an orbital angular momentum (OAM) of $l\hbar$ \cite{1, 2}, where $\varphi$ is the azimuthal angle and $l$ is the topological charge (TC) which can be any integer value. With the high dimensionality of the angular momentum, the OAM beams provide good sources both in classical and quantum optics for many different applications, such as optical tweezers and micromanipulation \cite{3,4,5,6,7}, classical optical communications \cite{8,9,10,10-1,10-2}, quantum cryptography \cite{qkd1,qkd2,qkd3}, high-dimensional quantum information \cite{11,12,13,14,14-1,14-2,14-3,14-4,14-5}, spiral phase contrast imaging \cite{15}, holographic ghost imaging \cite{16} and so on.
 
One of the crucial issues in these applications is the precise determination the TC of an unknown  OAM beam. Various methods have been  proposed to detect the TC of OAM beams. One general and convenient approach is through interference, for instance by interfering the measured OAM beam with a uniform plane wave or its mirror image \cite{17, 18}. Another good choice is the utilization of diffraction patterns with a special mask, such as triangular aperture diffraction \cite{19}, multiple-pinhole diffraction \cite{20}, single or double slit(s) diffraction \cite{21,22}, angular double slits diffraction \cite{23,24,24-1,24-2} and so on. Besides, geometric transformation by converting OAM state into transverse momentum states provides an efficient way  which converts OAM states into transverse momentum states\cite{25,26}. However, almost all of these methods strongly couple the measurement to the state resulting in the total destruction of the intensity distribution of the beams, or the OAM states.

An interesting experimental technique is to resort to weak measurement. First proposed by Aharonov \textit{et al.} \cite{27}, as an extension to the standard von Neumann model of quantum measurement, weak measurement is characterized by the pre- and post-selected states of the measured system. This approach extracts very little information about the measured system from a single outcome without causing the state collapse. This feature makes weak measurement an ideal tool for examining the fundamentals of quantum physics, such as for the measurement of the profile of the wave function \cite{28}, the realization of signal amplification \cite{29}, the resolution of the Hardy paradox \cite{30,31}, a direct measurement of density matrix of a quantum system \cite{32} and so on. Recently, a theoretical scheme is put forward to measure the OAM state through weak measurement process with the positive integral TC of the OAM beams \cite{33}. In this Letter, we experimentally demonstrate the method proposed in Ref \cite{33} and extend the value of TC $l$ to the negative $l$ regime.

Based on Ref. \cite{33}, with the value of  TC expending to the negative range  the unknown OAM state in the position space can be expressed as
\begin{align}
{ \langle x,y\vert\psi_{l}\rangle=\dfrac{\sqrt{2^{\vert l \vert+1}}}{\sigma\sqrt{\pi\vert l\vert
!}}(\dfrac{x+i{\rm sgn}(l)y}{\sigma})^{\vert l\vert}\exp(-\dfrac{x^{2}+y^{2}}{\sigma}^{2})}.
\label{Eq1}
\end{align}
where $l$ is the TC, and $ l\neq 0 $. Following the scheme of weak measurement detailed in Ref. \cite{33}, the initial state $  \vert \varphi\rangle_{i}$ is prepared as $  \vert \varphi \rangle_{i} = \vert i\rangle \bigotimes \vert\psi_{l}\rangle $, in which $ \vert i \rangle $ is the preselected state. By considering the von Neumann measurement, the interaction Hamiltonian can be described as $ \hat{H}=\gamma\hat{A}\bigotimes\hat{P} $, where $ \gamma $ is the coupling constant, $ \hat{A}$ is an observable of the preselected state and $ \hat{P}$ is the momentum observable of the unknown OAM state. Consequently, the unitary transformation is $ \hat{U}=e^{-i\gamma \hat{A}\bigotimes\hat{P}} $. After this unitary transformation, system $ \hat{A}$ is post-selected to the state $\vert f \rangle $ and the unknown OAM state is projected to $\vert x,y \rangle $ basis. Then the final state become $  \varphi(x,y) =\langle f\vert \langle x,y\vert \hat{U} \vert \varphi\rangle_{i} $ which contains the information  about $ l $. 

Under weak measurement $\gamma\ll\sigma $,so that $l^{2} \dfrac{\gamma^{2}}{\sigma^{2}}\ll 1 $, the measurement result is obtained as
\begin{align}
{ l=-\dfrac{\rm Re{\emph A_{w}}} {\rm Im \emph A_{w}}[\dfrac{\bar{y}}{\bar{x}}+O(l^{2} \dfrac{\gamma^{2}}{\sigma^{2}})]},
\label{Eq2}
\end{align}
Neglecting higher orders, we get
\begin{align}
{l\simeq l_{n}= 
-\dfrac{\rm Re  \emph A_{w}}{\rm Im \emph A_{w}}\dfrac{\bar{y}}{\bar{x}}},
\label{Eq2a}
\end{align}
where
\begin{align}
{ \bar{x}=\dfrac{\int x\varphi(x,y)^{*}\varphi(x,y)dxdy}{\int \varphi(x,y)^{*}\varphi(x,y)dxdy}},\nonumber\\
{ \bar{y}=\dfrac{\int y\varphi(x,y)^{*}\varphi(x,y)dxdy}{\int \varphi(x,y)^{*}\varphi(x,y)dxdy}},\nonumber
\end{align}
and
\begin{align}
{ A_{w}=\dfrac{\langle f\vert\hat{A}\vert i\rangle }{\langle f\vert i\rangle}}.\nonumber
\end{align}

\textsc{\begin{figure}[htb]
\centerline{\includegraphics[width=8.0 cm]{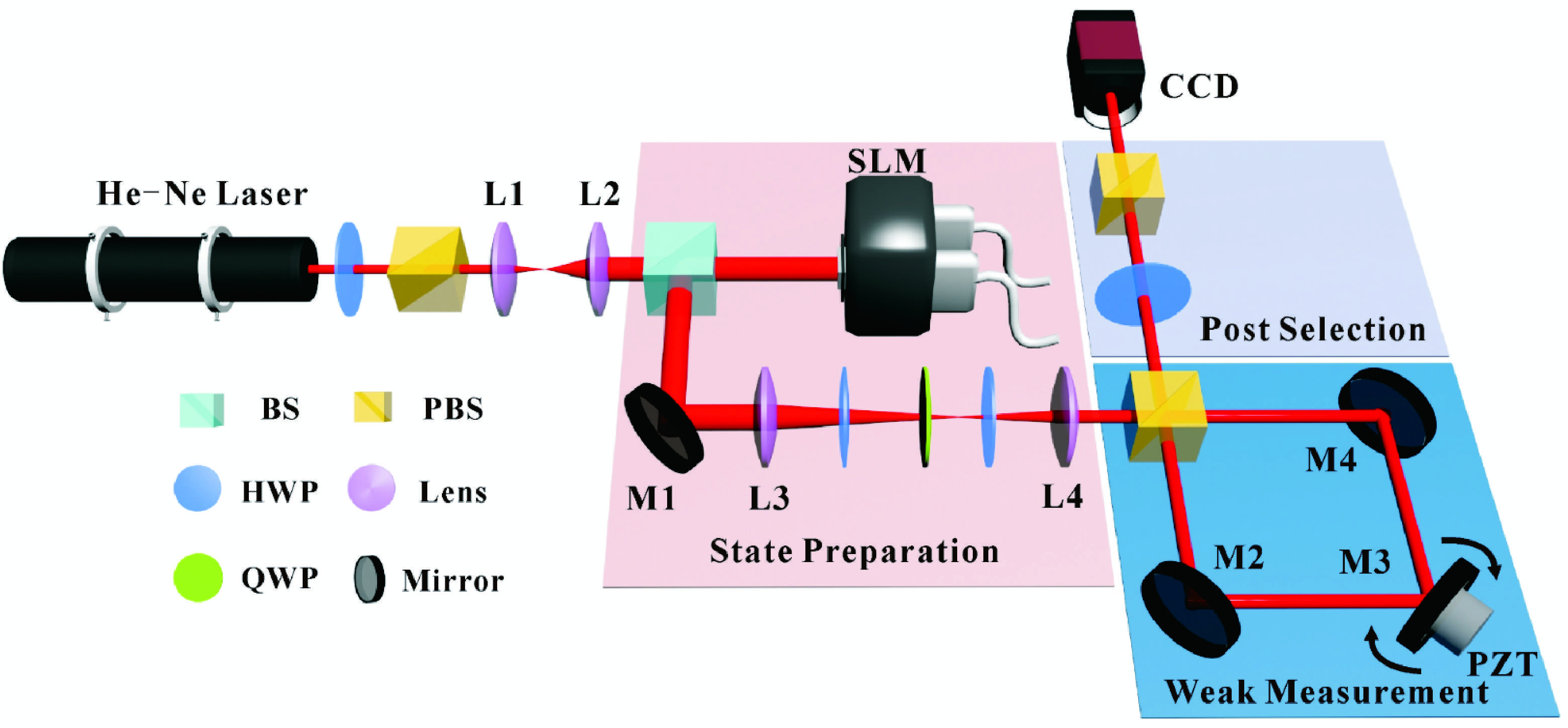}}
\caption{A sketch of the experimental setup. }
\end{figure}}

\textsc{\begin{figure*}[htb]
\centerline{\includegraphics[width=17.0 cm]{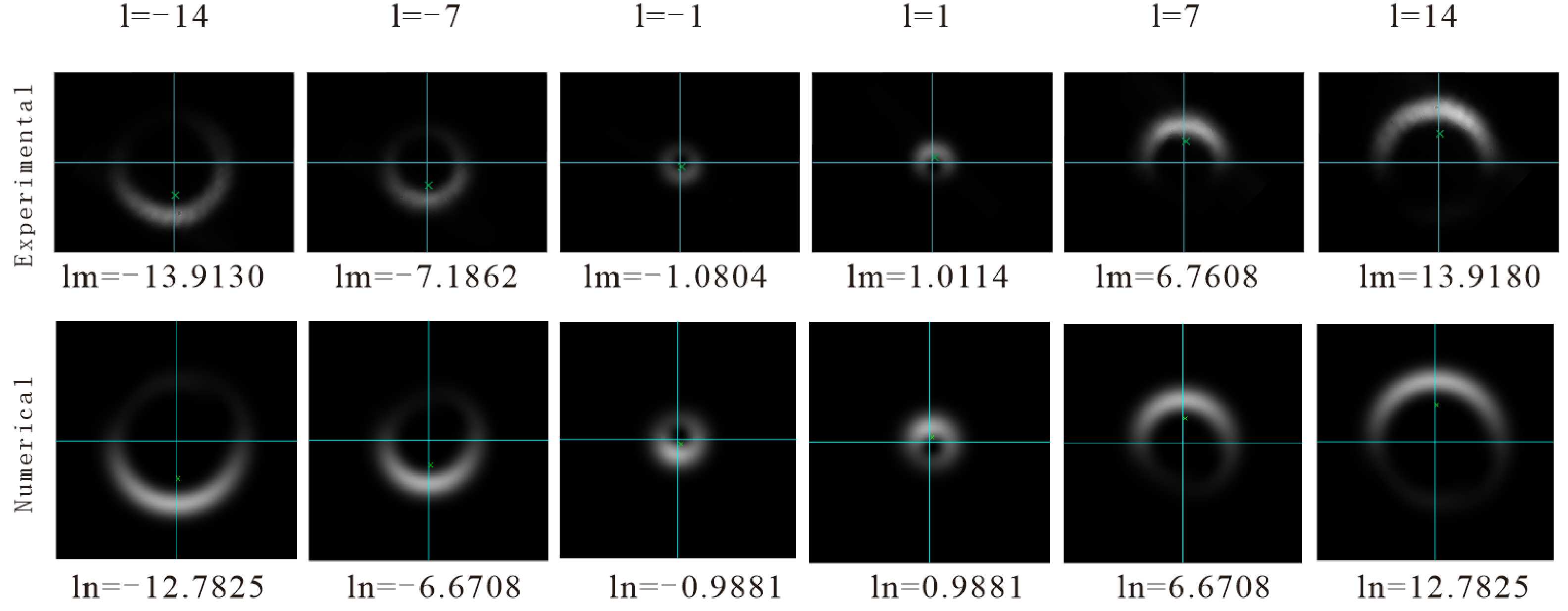}}
\caption{Intensity distributions of the final states for different TC of OAM beams. $l$ stands for the value of the OAM initially generated. $l_{m}$ is the experimentally measured TC of the beams and $l_{n}$ is the numerical result. The top row is the  intensity distribution of the final states in experiment while the bottom row is the simulative intensity distribution. The blue lines are the (x,y) coordinate lines and the green crosses denote centroids of the intensity distribution. }
\end{figure*}}

Figure 1 shows the experimental setup of the weak measurement. Light from a He-Ne laser passes through a half-wave plate (HWP) and a polarizing beam splitter (PBS). The HWP is used to adjust the light intensity and the PBS filters the desired polarization. The beam is then focussed by two lenses, L1 and L2. The focussed beam is vertically illuminated on a spatial light modulator (SLM) with the resolution of 20 $\mu$m per pixel to generate OAM states. The beam then passes through  two HWPs with a quarter-wave plate (QWP) inserted in-between to prepare a known polarization state as the preselected state $ \vert i \rangle $. At this point, the state preparation is completed. To achieve the weak measurement operation, a polarizing Sagnac interferometer is employed in which one of the  mirrors, M3, is connected with a piezo-transmitter (PZT). A single polarizing beam splitter (PBS) is used as the entry and exit gates of the device. When entering the interferometer, the incident beam is split  into different polarization components $ \vert H \rangle $ and  $ \vert V \rangle $, which traverse the interferometer in opposite directions. Without the PZT, the two components would combine again when they exit the PBS. Because of the existence of the PZT, a tiny  rotation can be imposed on M3  and the two components can be separated slightly in different directions at the exit. This executes exactly the operation $ \hat{U} $ we mentioned above. Meanwhile the lenses, L3 and L4, constitute the 4f system which images the distribution of both $H$ and $V$ components after the beam is reflected by the SLM to the position of the charge-coupled device (CCD) camera. After the weak measurement, a HWP and a PBS are used to post-select the state $ \vert f \rangle $. Finally, the intensity pattern is recorded by a CCD camera. 

In the setup used, the observable of the weak value $ \hat{A} $ corresponds exactly to $ \hat{A}=\vert H\rangle \langle H\vert-\vert V\rangle \langle V\vert $. The pre-selected state is prepared as  $ \vert i \rangle=a\exp(-i2\theta)\vert H \rangle+b\exp(i\phi)\exp(i2\theta)\vert V \rangle $ and the post-selected state is  $ \vert f \rangle=\cos(2\eta)\vert H \rangle+\sin(2\eta)\vert V \rangle $. The values of the parameters involved in our experiment are listed in Table 1.

\begin{table}[tbp]
\centering 
\caption{The values of experimental parameters}
\begin{tabular}{|c|c|c|c|c|c|} 
\hline
a & b & $\theta$ &$\phi$ &$\eta$ &$\dfrac{\gamma^{2}}{\sigma^{2}}$ \\ 
\hline  
0.7015 &0.7126 & $10^\circ$ &$17
0^\circ$ &$18.5^\circ$&0.006 \\         
 \hline
\end{tabular}

\end{table}

\begin{table*}[tbp]\tiny
\centering 
\caption{The TC values of experimental and numerical results. $l_{g}$, $l_{m}$ and $l$ are as the same as Fig. 2.  }
\begin{tabular}{|c|c|c|c|c|c|c|c|c|c|c|c|c|c|c|} 
\hline
$l_{g} $ & -1 & -2 &-3 &-4 &-5&-6 &-7 &-8 &-9 &-10 &-11 &-12 &-13 &-14 \\
\hline
$l_{m} $ &-1.0804 &-2.1603 &-3.0953 &-4.0695 &-5.0895 &-6.1386 &-7.1862 &-8.2243 &-9.2286 &-10.1509 &-11.3784       &-11.9755 & -13.3031 & -13.9130 \\
\hline

$l $ & -0.9881 &  -1.9644 &-2.9289 & -3.8817 &-4.8229 &-5.7526 &-6.6708 &-7.5775 &-8.4730 &-9.3571 &-10.2301 &-11.0919 &-11.9427 &-12.7825 \\
\hline
\hline
 $l_{g} $ & 1 & 2 &3 &4 &5&6 &7 &8 &9 &10 &11 &12 &13 &14 \\
\hline
$l_{m} $ & 1.0114 & 1.9817 &2.9786 &3.9707 &4.7446 & 5.7691 &6.7608 &7.9425 &8.8233 &10.0349 &11.1854 &12.357 &12.8498 &13.918 \\
\hline 

$l $ & 0.9881 & 1.9644 &2.9289 &3.8817 &4.8229 &5.7526 &6.6708 &7.5775 &8.4730 &9.3571 &10.2301 &11.0919 &11.9427 &12.7825 \\      
 \hline
\end{tabular}
\end{table*}

Figure 2 shows the experimental and the numerically simulated light intensity distribution of the final states.  Table 2 lists the experimental and simulated TC values of the OAM beams over a range $l=-14$ to $l=+14$. where $l$ represent the TC originally generated, $l_{m}$ is the one  experimentally measured and $l_{n}$ is gotten by numerically calculating. From Fig.1 and Tab. 2, it is clearly observed that the absolute value of experimental result is a little larger than the simulated one and the accuracy of the simulated value obviously decline more quickly than the experimental one with the increase of the $\vert l\vert$. This is caused by an experimental treatment. To avoid over exposure, we always adjust the parameters of the CCD to control the maximal incident light intensity. As a result, the intensity distribution recorded $I_{m}$ is the product of the actual distribution $I$ and a scaling coefficient $\mu$. For the  CCD has a response threshold so that the intensity below the value will not be recorded, the scale coefficient of the intensity will cause some intensity loss. Suppose that the actual light intensity is $I(x,y)=\varphi(x,y)^{*}\varphi(x,y)$ and the response threshold is $\xi I_{max}$, and then, the intensity we record in the picture ($I_{m}$) becomes 

\begin{equation}
 I_{m}(x,y)=\mu I(x,y)\Theta[\xi I_{max}-I(x,y)],
 \label{Eq2}
\end{equation}
where $I_{max}=\max[I(x,y)]$, and
\begin{equation}
 \Theta (t)=
 \begin{cases}
1 & t>0,\\
0 &t\leq 0.
\end{cases}\ 
\nonumber
\end{equation}
Replacing $\varphi(x,y)^{*}\varphi(x,y)$ in Eq. (3) with $I_{m}$, we get 

\begin{align}
{ l_{m}= 
-\dfrac{\rm Re  \emph A_{w}}{\rm Im \emph A_{w}}[\dfrac{\int y\varphi(x,y)^{*}\varphi(x,y)dxdy}{\int x\varphi(x,y)^{*}\varphi(x,y)dxdy-\int xO(\xi I(x,y))}} \nonumber\\
{-\dfrac{\int yO(\xi I(x,y))}{\int x\varphi(x,y)^{*}\varphi(x,y)dxdy-\int xO(\xi I(x,y))}]},
\label{Eq2}
\end{align}
where $O(\xi I(x,y))$ represents the effect of the faint intensity that is not recorded by the CCD. Since the second term of Eq.(5) is much smaller than the first term, we can ignore it. It is then obvious that $l_{m}$ is a little larger than $l_{n}$. For the Eq. (3) is derived by neglecting the higher terms of $\dfrac{\gamma^{2}}{\sigma^{2}}$, the result of $l_{n}$ is smaller than the value initially generated $l$. In  this sense,  Eq. (5) gives some compensation and the accuracy of the experimental results decrease more slowly. This means the restriction $l^{2} \dfrac{\gamma^{2}}{\sigma^{2}}\ll 1 $ in Ref. \cite{33} can be relaxed and the difficulty of the experiment can be reduced substantially.   In addition, all the theories involved are based on the fact that $l$ is an integer, so that it is reasonable to round $l_{m}$. In this situation, our experiment is perfectly consistent with the true value.

In conclusion, as an extension to the standard von Neumann model of quantum measurement weak measurement extracts very little information about the measured system and does not cause the measured state collapse. When coupling the OAM state with a well-defined pre- and post-selected system of a weak measurement process, there is an indirect coupling between $\bar{x}$, $\bar{y}$ and $l$. Based on this, we have proposed an effective scheme to  experimentally  measure the TCs of OAM  beams  by weak measurement.  In our experiment  the OAM value of the beams spanning from -14 to 14 are accurately probed. Note that since it is a weak measurement, the state does not collapse and it can be used for further experiments.

We thank Professor Changliang Ren for helpful discussions. This work is supported by the Fundamental Research Funds for the Central Universities, Joint Funds of the Ministry of Education of China (Grant No. 6141A02011604), Natural Science Foundation of Shaanxi Province (Grant No. 2017JM6011), and National Natural Science Foundation of China ( Grant Nos. 91736104, 11374008, and 11534008).


\begin{thebibliography}{99}
\bibitem{1}L. Allen, M. W. Beijersbergen, R. J. C. Spreeuw, and J. P. Woerdman, "Orbital angular momentum of light and the transformation of Laguerre-Gaussian laser modes," Phys. Rev. A \textbf{45}, 8185-8189 (1992).

\bibitem{2}A. M. Yao  and M. J. Padgett,"Orbital angular momentum:
origins, behavior and applications,"  Adv. Opt. Photon. \textbf{3}, 161-204 (2011).

\bibitem{3} H. He, M. E. J. Friese, N. R. Heckenberg, and H. Rubinsztein-Dunlop, "Direct observation of transfer of
angular momentum to absorptive particles from a laser
beam with a phase singul,"  Phys. Rev. Lett. \textbf{75}, 826-829 (1995).

\bibitem{4} K. Dholakia and T. \v{C}i\v{z}m\'{a}r, "Shaping the future of manipulation," Nature Photon. \textbf{5}, 335-342 (2011).



\bibitem{5} N. B. Simpson, K. Dholakia, L. Allen, and M. J. Padgett, "Mechanical equivalence of spin and orbital angular momentum of light : an optical spanner," Opt. Lett. \textbf{22}, 52-54 (1997).

\bibitem{6}D. G. Grier, "A revolution in optical manipulation," Nature \textbf{424}, 810-816 (2003).

\bibitem{7} M. J. Padgett and R. W. Bowman, "Tweezers with a
twist," Nature Photon. \textbf{5}, 343-348 (2011).

\bibitem{8} G. Gibson, J. Courtial, M. J. Padgett, M. Vasnetsov, V. Pas'ko, S. M. Barnett, and S. Franke-Arnold "Freespace
information transfer using light beams carrying orbital angular momentum,"  Opt. Express \textbf{12}, 5448-5465 (2004).

\bibitem{9} J. Wang, J. Y. Yang, I. M. Fazal, N. Ahmed, Y. Yan, H. Huang, Y. Ren, Y. Yue, S. Dolinar, M. Tur and A. E. Willner, "Terabit free-space data transmission employing
orbital angular momentum multiplexing," Nature Photon. \textbf{6}, 488-496 (2012). 

\bibitem{10} N. Bozinovic, Y. Yue, Y. Ren, M. Tur, P. Kristensen, H. Huang, A. E. Willner, and S. Ramachandran, "Terabit scale orbital angular momentum mode division multiplexing
in fibers," Science \textbf{340}, 1545-1548 (2013).

\bibitem{10-1}M. Krenn, J. Handsteiner, M. Fink, R. Fickler, R. Ursin, M. Malik and A. Zeilinger, "Twisted light transmission over 143 km," Proc. Natl. Acad. Sci. USA \textbf{113}, 13648-13653 (2016).

\bibitem{10-2} M.P. Lavery, C. Peuntinger, K. G\"{u}nthner, P. Banzer, D. Elser, R.W. Boyd, M.J. Padgett, C. Marquardt and G. Leuchs, "Free-space propagation of high-dimensional structured optical fields in an urban environment," Science Advances \textbf{3}, e1700552 (2017).

\bibitem{qkd1} S. Gr\"{o}blacher, T. Jennewein, A. Vaziri, G. Weihs and A. Zeilinger, "Experimental quantum cryptography with qutrits," New J. Phys. \textbf{8}, 75 (2006).

\bibitem{qkd2} M. Mirhosseini, O.S. Maga\~{n}a-Loaiza, M.N. O'Sullivan, B. Rodenburg, M. Malik, M.P. Lavery, M.J. Padgett, D.J. Gauthier and R.W. Boyd, "High-dimensional quantum cryptography with twisted ligh," New J. Phys. \textbf{17}, 033033 (2015).

\bibitem{qkd3} A. Sit, F. Bouchard, R. Fickler, J. Gagnon-Bischoff, H. Larocque, K. Heshami, D. Elser, C. Peuntinger, K. G\"unthner, B. Heim, C. Marquardt, G. Leuchs, R. W. Boyd, and E. Karimi, "High-dimensional intracity quantum cryptography with structured photons," Optica \textbf{4}, 1006-1010 (2017).

\bibitem{11} A. Mair, A. Vaziri, G. Weihs and A. Zeilinger, "Entanglement of the orbital angular momentum states of photons,"  Nature \textbf{412}, 313-316 (2001). 

\bibitem{12} G. Molina-Terriza, J. P. Torres, and L. Torner, "Management of the Angular Momentum of Light: Preparation of Photons in Multidimensional Vector States of Angular Momentum," Phys. Rev. Lett. \textbf{88}, 013601 (2002).

\bibitem{13} A. Vaziri, G. Weihs and A. Zeilinger, "Experimental Two-Photon, Three-Dimensional Entanglement for Quantum Communication," Phys. Rev. Lett. \textbf{89}, 240401 (2002).

\bibitem{14} J. E. Curtis and D. G. Grier, "Structure of Optical Vortices," Phys. Rev. Lett. \textbf{90}, 133901(1) (2003).

\bibitem{14-1} J. Leach, B. Jack, J. Romero, A. K. Jha, A. M. Yao, S. Franke-Arnold, D. G. Ireland, R. W. Boyd, S. M. Barnett, and M. J. Padgett, "Quantum Correlations in Optical Angle–Orbital Angular Momentum Variables," Science \textbf{329}, 662-665 (2010).

\bibitem{14-2} A.C. Dada, J. Leach, G.S. Buller, M.J. Padgett and E. Andersson, "Experimental high-dimensional two-photon entanglement and violations of generalized Bell inequalities," Nature Physics \textbf{7}, 677-680 (2011). 

\bibitem{14-3} J. Romero, D. Giovannini, S. Franke-Arnold, S. Barnett and M. Padgett, "Increasing the dimension in high-dimensional two-photon orbital angular momentum entanglement," Phys. Rev. A \textbf{86}, 012334 (2012).

\bibitem{14-4} X.-L. Wang, X.-D. Cai, Z.-E. Su, M-C. Chen, D. Wu, L. Li, N.-L. Liu, C.-Y. Lu, and J.-W. Pan, "Quantum teleportation of multiple degrees of freedom of a single photon,"  Nature \textbf{518}, 516-519 (2015).

\bibitem{14-5} D.-X. Chen, R.-F. Liu, P. Zhang, Y.-L. Wang, H.-R. Li, H. Gao, F.-L. Li, "Realization of quantum permutation algorithm in high dimensional Hilbert space," Chin. Phys. B \textbf{26}, 060305 (2017).

\bibitem{15} S. F\" urhapter, A. Jesacher, S. Bernet, and M. Ritsch-Marte, "Spiral phase contrast imaging in microscopy,"  Opt. Express \textbf{13}, 689-694 (2005).

\bibitem{16} B. Jack, J. Leach, J. Romero, S. Franke-Arnold, M. Ritsch-Marte, S. M. Barnett, and M. J. Padgett, ""Holographic
ghost imaging and the violation of a bell inequality," Phys. Rev. Lett. \textbf{103}, 083602 (2009). 

\bibitem{17}M. Harris, C. A. Hill, P. R. Tapster and J. M. Vaughan,  "Laser modes with helical wave fronts," Phys. Rev. A \textbf{49}, 3119 (1994).

\bibitem{18}M. J. Padgett, J. Arlt, N. B. Simpon and L. Allen, "An experiment to observe the intensity and phase structure of Laguerre Gaussian laser modes,"  Am. J.Phys. \textbf{64}, 77-82 (1996).

\bibitem{19}A. Mourka, J. Baumgartl, C. Shanor, K. Dholakia, and E. M. Wright, "Visualization of the birth of an optical vortex using diffraction from a triangular aperture," Opt. Express \textbf{19}, 5760-5771 (2011).

\bibitem{20} G. C. G. Berkhout and M. W. Beijersbergen, "Method
for Probing the Orbital Angular Momentum of Optical Vortices in Electromagnetic Waves from Astronomical Objects,"  Phys. Rev. Lett. \textbf{101}, 100801 (2008).

\bibitem{21} H. I. Sztul and R. R. Alfano, "Double-slit interference
with Laguerre-Gaussian beams,"  Opt. Lett. \textbf{31}, 999-1001 (2006).

\bibitem{22} Q. S. Ferreira, A. J. Jesus-Silva, E. J. S. Fonseca  and  J. M. Hickmann, "Fraunhofer diffraction of light with
orbital angular momentum by a slit,"  Opt. Lett. \textbf{36}, 3106-3108 (2011).

\bibitem{23} H. Zhou, L. Shi, X. Zhang, and J. Dong, "Dynamic interferometry measurement of orbital angular momentum of
light,"  Opt. Lett. \textbf{39}, 6058-6061 (2014).

\bibitem{24} D. Fu, D. Chen, R. Liu, Y. Wang, H. Gao, F. Li, and P. Zhang, "Probing the topological charge of a vortex
beam with dynamic angular double slits," Opt. Lett. \textbf{40}, 788-791 (2015).

\bibitem{24-1} J. Zhu,  P. Zhang, D. Fu, D. Chen, R. Liu, Y. Zhou, H. Gao, and F. Li, "Probing the fractional topological charge of a vortex light beam by using dynamic angular double slits," Photonics Research \textbf{4}, 187-190 (2016).

\bibitem{24-2} J. Zhu, P. Zhang, D. Chen, R. Liu, Y. Zhou, J. Wang, H. Gao, and F. Li, "Robust method to probe the topological charge of
a Bessel beam by dynamic angular double slits," Appl. Opt. \textbf{57}, B39-B44 (2018).

\bibitem{25}G. C. G. Berkhout, M. P. J. Lavery, J. Courtial, M. W. Beijersbergen, and M. J. Padgett, "Effcient sorting of orbital
angular momentum states of light," Phys. Rev. Lett. \textbf{105}, 153601 (2010).

\bibitem{26} M. Mirhosseini, M. Malik, Z. Shi and R. W. Boyd, "Efficient separation of the orbital angular momentum eigenstates
of light,"  Nat. Commun. \textbf{4}, 2781 (2013).

\bibitem{27}Y. Aharonov, D. Z. Albert, and L. Vaidman, "How the
result of a measurement of a component of the spin of a spin-1/2 particle can turn out to be 100," Phys. Rev. Lett. \textbf{60}, 1351-1354 (1988).

\bibitem{28}J. S. Lundeen, B. Sutherland, A. Patel, C. Stewart, and C.Bamber, "Direct measurement of the quantum wavefunction,"  Nature \textbf{474}, 188-191 (2011).

\bibitem{29}O. Hosten and P. Kwiat, "Observing the Spin Hall Effect
of Light via Quantum Weak Measurements,"  Science \textbf{319}, 787-790 (2008).

\bibitem{30}Y. Aharonov, A. Botero, S.  Popescu, B. Reznik, and J. Tollaksen, "Revisiting Hardy's paradox: counterfactual statements, real measurements, entanglement and weak values,"  Phys. Lett. A \textbf{301}, 130-138 (2002).

\bibitem{31}J. S. Lundeen and A. M. Steinberg, "Experimental Joint
Weak Measurement on a Photon Pair as a Probe of
Hardy's Paradox,"  Phys. Rev. Lett. \textbf{102}, 020404 (2009).

\bibitem{32} G. S. Thekkadath, L. Giner, Y. Chalich, M. J. Horton, J. Banker, and J. S. Lundeen, "Direct Measurement of the Density Matrix of a Quantum System," Phys. Rev. Lett. \textbf{117}, 120401 (2016). 

\bibitem{33}J. Qiu, C. Ren, and Z. Zhang, "Precisely measuring the
orbital angular momentum of beams via weak measurement," Phys. Rev. A \textbf{93}, 063841 (2016).

\end{thebibliography}
\end{document}